\documentclass{aa}
\usepackage[varg]{txfonts}
\usepackage{array} 
\usepackage{graphicx}

\usepackage{txfonts}

\usepackage{natbib}
\begin{document}

   \title{\bf A data base of synthetic photometry in the GALEX ultraviolet bands for the stellar sources observed with the International Ultraviolet Explorer}

   \subtitle{}

   \author{Leire Beitia-Antero
          \inst{1}
          \and
          Ana I. G\'omez de Castro \inst{1}
          }

   \institute{AEGORA Research Group, Universidad Complutense de Madrid,
	Facultad de CC. Matem\'aticas, Plaza de Ciencias 3, E-28040 Madrid\\
              \email{aig@ucm.es}
             }

   \date{Received November 17th, 2015; accepted}

\titlerunning{UV photometric data base}
\authorrunning{Beitia-Antero and G\'omez de Castro}
 
  \abstract
   {The  Galaxy Evolution Explorer (GALEX) has produced the largest photometric catalogue of ultraviolet (UV) sources. As such, it has defined the 
new standard bands for UV photometry: the near UV band (NUV) and the far UV band (FUV). However, due to brightness limits, the GALEX mission has avoided
the  Galactic plane which is crucial for astrophysical research and future space missions.  }
   {The International Ultraviolet Explorer (IUE) satellite obtained 63,755 spectra in the low dispersion mode ($\lambda/\delta \lambda \sim 300$ )  during its 18 years lifetime. We have derived the photometry in the GALEX bands for all the stellar sources in the IUE Archive to extend the GALEX data base
with observations including the Galactic plane.}
   {Good quality spectra have been selected for the IUE classes of stellar sources. The GALEX FUV and NUV magnitudes
have been computed using the GALEX transmission curves, as well as the conversion equations between flux and magnitudes
provided by the mission (galexgi.gsfc.nasa.gov).}
   { Consistency between GALEX and IUE synthetic photometries has been tested using White Dwarfs (WD) contained in 
both samples. The  non-linear response performance of GALEX inferred from this data agrees with the 
results from GALEX calibration. The photometric data base  is made available to the community through the services of the 
Centre de Données Stellaires at  Strasbourg (CDS). The catalogue contains  FUV magnitudes for 1,631 sources, ranging from FUV= 1.81 to FUV=18.65 mag. In the NUV band, the catalogue
includes observations for  1,005 stars ranging from NUV = 3.08 to NUV= 17.74 mag . }
{UV photometry for  1,493 not included in the GALEX AIS GR5 catalogue is provided; most of them are hot (O-A spectral type) stars.
The sources in the catalogue are distributed over the full sky, including the Galactic plane.}


   \maketitle
%

\section{Introduction}

The catalogue of ultraviolet (UV) sources generated by the Galaxy Evolution Explorer
(GALEX) mission constitutes the most extensive data base of UV photometry (Martin et al. 2005, hereafter Ma2005;
Bianchi 2014, hereafter B2014). As such,  the GALEX near (UV) or NUV band and far (UV) or FUV band have become standards for
the description of the Spectral Energy Distribution (SED) of sources in broad band photometry.
The NUV band ranges from 1771~\AA\  to 2831~\AA\ with effective wavelength 2315.7~\AA\ and 
the FUV band ranges from 1344~\AA\ to 1786 ~\AA\ with effective wavelength 1538.6~\AA\  
(Morrissey et al. 2007;  hereafter M2007).

The UV detectors used by the GALEX mission were sensitive MCP type detectors
with a global count rate limit of 100,000 counts~s$^{-1}$ (M2007). As a result, the GALEX survey
avoided the Galactic plane (B2014). Moreover, the photometric data base does not extend to bright UV sources
that are ideally suited for calibration purposes.

The International Ultraviolet Explorer (IUE) (Boggess et al. 1978;  herefater B1978) contains the largest data set of UV spectra.
Most of them were obtained in photometric conditions: good guiding, large aperture  (10x20 arcsec) and
low dispersion ( $\lambda/\delta \lambda \sim 300$). We have used this spectral data base to compute the FUV and NUV synthetic magnitudes of all 
stellar sources observed with IUE under this configuration (31,982 spectra).

In this research note, we describe the methods followed to derive the FUV and NUV magnitudes from the
IUE spectra and quantify the photometric accuracy of these results by comparing GALEX and 
IUE-based photometry for the white dwarfs (WDs) contained in both samples. We also describe the contents and 
characteristics of the data base submitted to the Centre de Données Stellaires at Strasbourg (CDS).

\section{The IUE data base of stellar spectra}

The IUE data base contains 63,755 spectra obtained through  the large aperture  (10x20 arcsec)  in low dispersion
mode, from those we have selected only the  stellar 
sources\footnote{ \url{http://sdc.cab.inta-csic.es/ines/InForm.html#class}}.  
This amounts to a grand total of 31,982 stellar spectra\footnote{No included in this census are:
IUE sky images misclassified among the stellar sources or individual objects belonging to the SMC, LMC, clusters and 
associations with identification no recognized by the Centre de Donn\'ees Stellaires (CDS, Strasbourg).}.\\

The IUE low dispersion spectra were recorded  with three cameras: long wavelength prime (LWP), long wavelength
redundant (LWR) and short wavelength prime (SWP)  (B1978). From the 31,982 stellar spectra, 16,467  are \texttt{SWP} observations, 
10,349 \texttt{LWP} observations and 5,166 \texttt{LWR} observations.  Good 
quality spectra (according to the criteria defined in Sect.~4) are available for 1889 stars in the SWP camera and 1157 stars in the LWP or LWR cameras.

\section{The GALEX photometric bands in the IUE spectra}

GALEX was a 50-cm primary space telescope with a Ritchey-Chr\'etien mounting feeding simultaneously two detectors 
sensitive to the near and far  UV  making use of a multilayer dichroic beamsplitter (Morrissey et al. 2005,  Ma2005).
The GALEX photometric bands are defined in the mission documentation (M2007)
and can be downloaded from the \emph{GALEX} 
official webpage\footnote{\url{http://galexgi.gsfc.nasa.gov/docs/galex/Documents/PostLaunchResponseCurveData.html}}. 
The FUV and NUV bands cover the spectral ranges 1344-1786~\AA\ and 1771-2831~\AA\ respectively (M2007); 
the transmittance curves are shown in  Figure~1. FUV and NUV AB magnitudes are determined by means of the conversion:

\noindent
\begin{equation}
\text{FUV} = -2.5 \times \log \bigg( \frac{\text{Flux \, FUV}}{1.40 \times 10^{-15} \,
{\text{erg} \,\text{s}}^{-1}{\text{cm}}^{-2} {\AA}^{-1}} \bigg) + 18.82 
\label{eq:FUV}
\end{equation}

\begin{equation}
\text{NUV} = -2.5 \times \log \bigg( \frac{\text{Flux \, NUV}}{2.06 \times 10^{-16} \,
{\text{erg} \,\text{s}}^{-1}{\text{cm}}^{-2} {\AA}^{-1}} \bigg) + 20.08 
\label{eq:NUV}
\end{equation}

\noindent
(M2007).\\

The IUE long (1850 - 3300~\AA ) and short (1150 - 1980~\AA ) wavelength ranges 
(B1978) do  not match exactly the GALEX bands. 
While GALEX FUV is completely contained in the  IUE SW range, 
GALEX NUV is split between the LW and SW cameras. For this reason,  we 
need to join SWP and LW spectra in order to compute NUV magnitudes. 
Therefore NUV photometry can only be provided for non-variable sources.  

\begin{figure}[h!]
  \centering
  \includegraphics[width = 80 mm]{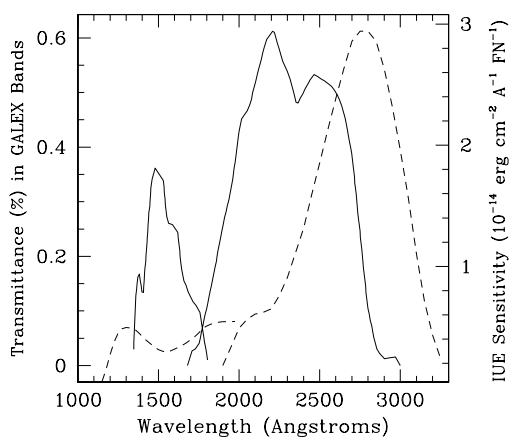}
  \caption{  GALEX transmission curves (solid) compared with IUE sensitivity curves (dashed)
for the SWP and LWR cameras from Bohlin et al. 1980. }
  \label{fig1}
\end{figure}

\section{Methods}

\subsection{Data Selection}

Spectra have been retrieved from the IUE Newly Extracted Spectra (INES) Archive; data in this archive 
were processed using an optimised extraction algorithm that prevents the introduction of artifacts during the 
extraction while preserving the photometric accuracy (see Rodr\'iguez-Pascual et al. 1999, 
for more details). The INES Archive is a final release from the European IUE Observatory.

The spectra are provided as \texttt{FITS} files binary
tables that  include four columns with: \textit{Wavelength} (\AA), \textit{Flux} ($erg ~ cm^{-2}~ s^{-1} ~ \text{\AA}^{-1}$), 
\textit{Flux error} ($erg ~ cm^{-2} ~ s^{-1} ~ \text{\AA}^{-1}$)   after pipeline processing\footnote{
(see Gonzalez-Riestra, Cassatella, Wamsteker 2001 for the absolute calibration of the INES IUE fluxes)}  and 
\textit{Quality Flag} (QF), a quality measure for each pixel.   $QF=0$ indicates that the pixel flux is reliable,
otherwise $QF <0$. Negative QFs are given in powers of $2$ ($QF =-2^n$ with $n = 1 \cdots14$)  
and $n$  is an identifier of  the various possible sources of inaccuracies in the pixel photometry:
microphonic noise, saturated pixel, reseaux mark, etc. (see \url{https://archive.stci.eduiue/manual/newsips/node20.html} for details). 

The data base contains {\it all} data including overexposed, sub-exposed spectra as well as 
observations that suffered problems during the downlink, calibration, 
processing, etc. 
To avoid including bad spectra in the photometric data base, we removed from our list:

\begin{itemize}
\item spectra with more than a 10\% of the pixels in the IUE spectrum flagged with  $QF < 0$. The vast majority of the spectra
contain reseau marks for the geometric calibration of the raw data and pixels flagged with $QF<0$ close  to the edge of 
the spectrum.  For this reason we have worked with a  tolerance of a 10\%. 

\item spectra with average \textit{Flux} smaller than three times the average \textit{Flux error}  in the corresponding GALEX wavelength range. 
Calibrated data include flux and flux error  at each  wavelength; this is the standard $3 \sigma$ criterium applied to 
remove sub exposed spectra from the working list.
\end{itemize}

LW spectra  require a more detailed examination since the peak transmittance of the GALEX NUV filter is at 2300~\AA ,
in an area  often sub-exposed in the spectrum of the cool stars observed with the IUE.
To evaluate the SNR in this region, we define four windows in the 1975~\AA\ - 2375~\AA\ range:
\texttt{REG~I} ($1975-2075 \text{\AA}$), \texttt{REG~II} ($2075-2175 \text{\AA}$), 
\texttt{REG~III} ($2175-2275 \text{\AA}$) and 
\texttt{REG~IV} ($2275-2375 \text{\AA}$). Within these windows we compute the 
mean flux and the dispersion.
We reject the spectra if the mean flux is negative  or if the standard deviation 
is ten times higher than the mean in any of the regions \texttt{II-III-IV}. We allow 
for a factor of ten to prevent removing sources with steep energy distributions
or strong features.

\subsection{Calculation of the FUV and NUV synthetic magnitudes}\label{sec:Calc_FUV_NUV}

The integrated flux in the FUV band is computed from the SWP IUE
spectra after multiplication by the normalised 
transmittance of the FUV filter. The flux of the flagged bad
pixels ($QF < 0$) has been interpolated from good nearby pixels.  The AB 
magnitude is calculated using a  Eq.~1 (see Sect.~3).

To evaluate the NUV synthetic magnitude is necessary to determine firstly,
whether the sources are variable (see Sect.~4.4 for a description of the procedure). 
In case there is not evidence of variability, 
the SW and LW are joined into a single spectrum; the matching wavelength
is set at 1975~\AA . After, the spectrum is multiplied by the normalised 
transmittance of the NUV filter.   The flux of the flagged bad
pixels ($QF < 0)$ has been interpolated from good nearby pixels.  The AB 
magnitude is calculated using a  Eq.~2 (see Sect.~.3). 

In case there are multiple observations and the source is found not to
vary, synthetic  magnitudes are computed from the average flux.

\subsection{Error determination}

For each spectrum and band, the \textit{Flux error} provided by the mission
(column \#3 in the data) is multiplied by the GALEX normalised transmittance to evaluate the total error in each band
denoted as $S_{FUV}$ and $S_{NUV}$  for the FUV and NUV bands respectively.
From that, magnitude errors are provided for the FUV band, eFUV$^+$ and eFUV$^-$, and 
NUV band, eNUV$^+$ and eNUV$^-$; note that the errors are
asymmetric around the magnitude value because of the logarithmic scale.

In case there are multiple observations and the source is found not to
vary, errors are from the \textit{Flux errors}.

\subsection{Multiple observations. Variable stars}\label{sec:variable_stars}

There are 598 stars in the IUE archive with multiple observations in the SW range and 
409 stars with multiple observations in the LW range.  For those variability has been tested.

\subsubsection{Stars with multiple observations in the IUE SW range}

We compute for each SWP observation, $i$,  the weighted flux in the GALEX FUV band, 
$F_{FUV}(i)$, calculated as:

\noindent
\begin{equation}
F_{FUV}(i) =\int  F_i(\lambda)G_{FUV}(\lambda)d\lambda
\label{eq:FG}
\end{equation}
\noindent
with $G_{FUV}$ being the normalized GALEX FUV transmission curve (see Figure~1). 
The weighted {\it Flux error},  $S_{FUV}$(i), is determined in the same manner: 
\noindent
\begin{equation}
S_{FUV}(i) =\int  S_i(\lambda)G_{FUV}(\lambda)d\lambda
\label{eq:sFG}
\end{equation}
\noindent
with $S_{FUV}(i)$ being the  {\it Flux error} (see Figure~1). 

After, the average,  $\langle F_{FUV} \rangle $,  the dispersion, $\sigma(F_{FUV})$, and the
average {\it Flux error}, $\langle S_{FUV} \rangle$ are computed.
We flag a star as variable if $\sigma(F_{FUV}) \ge 3 \times \langle S_{FUV} \rangle$.
Note that very noisy data have already been rejected in the data selection process (see Sect.~4.1).

\subsubsection{Stars with multiple observations in the IUE LW range}

The procedure is similar to the described for the SW range but, in this case,
the variability test is carried out only over the range of the NUV band contained
in the LW images. 

We  compute for each LW observation, $i$,  the weighted flux in the 1975 - 3000~\AA\ range, 
$F_{GaLW}(i)$, calculated as:
\noindent
\begin{equation}
F_{GaLW}(i) =\int _{1975} ^{3000} F_i(\lambda)G_{NUV}(\lambda)d\lambda
\label{eq:NG}
\end{equation}
\noindent
with $G_{NUV}$ being the normalized GALEX NUV transmission curve (see Figure~1), 
as well as the weighted {\it Flux error}, $S_{GaLW}$(i),
\noindent
\begin{equation}
S_{GaLW}(i) =\int _{1975} ^{3000} S_i(\lambda)G_{NUV}(\lambda)d\lambda
\label{eq:sNG}
\end{equation}
\noindent

After, the average,  $\langle F_{GaLW} \rangle $,  the dispersion, $\sigma(F_{GaLW})$, and the
average {\it Flux error}, $\langle S_{GaLW} \rangle$ are computed.
We flag a star as variable if $\sigma(F_{GaLW}) \ge 3 \times \langle S_{GaLW} \rangle$.

\subsubsection{In summary}

According to these criteria 52 stars are found to be variable in the LW range and 89 in the SW range;
only 36 stars are found to be variable in both ranges.   Therefore, $\sim 13$\% of the stars with multiple 
observations in the LW range are found to be variable and $\sim 15$\% of the stars with multiple 
observations in the SW range are found to be variable.

\section{Photometric accuracy}

The IUE sample (of non variable stars) provides good synthetic
photometry for 103 WDs. 43 of these 103 WDs have a 
counterpart in the GALEX GR5 AIS survey (Bianchi et al. 2011) within a 
search radius of  3''.
We have used this subset to check the photometric accuracy of the synthetic magnitudes
computed in this work (see Table~1 for their synthetic and GALEX magnitudes). 
The limiting magnitude of GALEX AIS\footnote{\url{https://archive.stsci.edu/prepds/gcat/gcat_gasc_gmsc.html}} 
(NUV $\sim$20.5~mag) is high above the sensitivity threshold of IUE in 
low dispersion mode, therefore counterparts are identified for all sources within the area mapped by GALEX.
Note that in May 2009, the FUV detector in GALEX  stopped working and as a result, releases later to GR5 (such as GR6/7) only 
add new sources in the NUV band\footnote{In fact, late GALEX releases also add very weak FUV sources identified by 
being the counterpart of a NUV source. These sources are too weak to have been observed with IUE.} (B2014). 

\begin{table*}[]
  \centering
  \caption{Synthetic photometry in the GALEX bands of WD observed with IUE: WDs with GALEX counterpart} 
  \begin{scriptsize}
  \begin{tabular}{lllllllllllll}
    
    \hline
    ID IUE & RAS (ICRS) & DEC (ICRS) & \multicolumn{4}{c}{IUE based synthetic photometry} & \multicolumn{4}{c}{GALEX  photometry} \\
              &                   &                  &  FUV & +eFUV & -eFUV & NUV & +eFUV & -eFUV & FUV & eFUV & NUV & eNUV  \\
    \hline

WD 0005+511 & 00:08:18.17 & +51:23:16.6 & 11.09 & 0.06 & -0.06 & NA & NA & NA & 12.190 & 0.004 & 13.178 & 0.004 \\
EQ 0027+259 & 00:29:37.96 & +26:10:28.47 & 13.60 & 0.07 & -0.06 & 13.22 & 0.30 & -0.24 & 13.567 & 0.008 & 13.785 & 0.006 \\
WD 0039+04 & 00:42:6.12 & +05:09:23.36 & 12.01 & 0.08 & -0.07 & NA & NA & NA & NA & NA & 13.189 & 0.004 \\
WD 0050-332 & 00:53:17.44 & -32:59:56.6 & 11.47 & 0.07 & -0.07 & 12.13 & 0.11 & -0.10 & NA & NA & 12.365 & 0.003 \\
WD 0112+104 & 01:14:37.8 & +10:41:6 & 14.40 & 0.15 & -0.13 & 14.66 & 0.28 & -0.22 & 14.425 & 0.010 & 14.623 & 0.007 \\
WD 0134+833 & 01:41:28.74 & +83:34:58.9 & 12.34 & 0.07 & -0.06 & 12.70 & 0.10 & -0.09 & 12.923 & 0.002 & 12.938 & 0.002 \\
WD 0232+035 & 02:35:7.59 & +03:43:56.8 & 10.39 & 0.07 & -0.06 & 11.01 & 0.11 & -0.10 & NA & NA & 12.371 & 0.003 \\
WD 0302+027 & 03:04:37.34 & +02:56:57.9 & 13.31 & 0.11 & -0.10 & 13.69 & 0.13 & -0.11 & 13.008 & 0.007 & 13.968 & 0.007 \\
WD 0320-539 & 03:22:14.83 & -53:45:16.5 & 13.13 & 0.09 & -0.08 & NA & NA & NA & 13.325 & 0.004 & 13.747 & 0.003 \\
WD 0342+026 & 03:45:34.58 & +02:47:52.81 & 10.44 & 0.07 & -0.07 & 10.76 & 0.12 & -0.11 & NA & NA & 12.751 & 0.003 \\
WD 0453-296 & 04:55:35.94 & -29:28:59.96 & 15.01 & 0.21 & -0.17 & 15.03 & 0.23 & -0.19 & 15.029 & 0.015 & 14.995 & 0.010 \\
HS 0713+3958 & 07:17:2.7 & +39:53:25 & 14.57 & 0.11 & -0.10 & 15.05 & 0.23 & -0.19 & 14.643 & 0.012 & 14.977 & 0.009 \\
WD 0846+249 & 08:49:5.88 & +24:45:7.93 & 14.45 & 0.08 & -0.08 & NA & NA & NA & 14.436 & 0.013 & 15.147 & 0.011 \\
WD 0853+163 & 08:56:18.96 & +16:11:3.8 & 15.28 & 0.14 & -0.13 & 15.31 & 0.50 & -0.34 & 15.273 & 0.022 & 15.201 & 0.013 \\
WD 0945+246 & 09:48:46.65 & +24:21:26 & 14.38 & 0.08 & -0.08 & 13.89 & 0.21 & -0.18 & 14.479 & 0.011 & 13.955 & 0.005 \\
PG 0958-073 & 10:00:47.25 & -07:33:31.01 & 12.44 & 0.07 & -0.06 & 12.90 & 0.13 & -0.11 & 12.517 & 0.006 & 13.190 & 0.005 \\
WD 1034+001 & 10:37:3.81 & -00:08:19.31 & 10.85 & 0.06 & -0.06 & 11.62 & 0.09 & -0.09 & NA & NA & 12.865 & 0.002 \\
WD 1114+072 & 11:16:49.35 & +06:59:33 & 11.85 & 0.07 & -0.06 & 12.24 & 0.13 & -0.11 & 12.360 & 0.003 & 13.119 & 0.004 \\
WD 1144+005 & 11:46:35.17 & +00:12:33.6 & 13.69 & 0.09 & -0.09 & NA & NA & NA & 12.921 & 0.008 & 13.691 & 0.006 \\
WD 1211+332 & 12:13:56.25 & +32:56:31.4 & 12.64 & 0.07 & -0.06 & 13.22 & 0.11 & -0.10 & 12.718 & 0.008 & 13.396 & 0.004 \\
WD 1230+052 & 12:33:12.57 & +04:57:37.7 & 12.04 & 0.07 & -0.06 & 12.34 & 0.10 & -0.09 & 12.544 & 0.004 & 13.732 & 0.004 \\
WD 1233+427 & 12:35:51.14 & +42:22:39.72 & 10.82 & 0.06 & -0.06 & 11.13 & 0.12 & -0.11 & 12.710 & 0.007 & 12.831 & 0.004 \\
WD 1302+283 & 13:04:48.63 & +28:07:29.3 & 13.69 & 0.11 & -0.10 & 14.10 & 0.24 & -0.19 & 13.802 & 0.009 & 14.238 & 0.005 \\
WD 1326-037 & 13:29:16.39 & -03:58:51.75 & 15.22 & 0.12 & -0.11 & 15.20 & 0.33 & -0.26 & 15.295 & 0.015 & 15.204 & 0.009 \\
WD 1406+590 & 14:08:32.23 & +59:40:25.1 & 11.93 & 0.11 & -0.10 & 12.28 & 0.21 & -0.18 & 12.690 & 0.003 & 12.548 & 0.002 \\
WD 1445+152 & 14:48:14.38 & +15:04:49.93 & 15.62 & 0.29 & -0.23 & 15.57 & 0.38 & -0.28 & 15.462 & 0.027 & 15.421 & 0.016 \\
WD 1636+351 & 16:38:26.32 & +35:00:11.9 & 13.09 & 0.09 & -0.09 & 13.75 & 0.37 & -0.28 & 13.236 & 0.005 & 13.854 & 0.003 \\
PG 1637+346 & 16:39:36.02 & +34:32:30.36 & 15.01 & 0.18 & -0.16 & NA & NA & NA & 14.403 & 0.008 & 14.656 & 0.005 \\
HS 1650+7229 & 16:49:16.1 & +72:24:12 & 15.44 & 0.12 & -0.11 & NA & NA & NA & 15.581 & 0.032 & 16.050 & 0.013 \\
WD 1708+602 & 17:09:15.9 & +60:10:11 & 11.82 & 0.04 & -0.04 & NA & NA & NA & 12.707 & 0.005 & 12.565 & 0.003 \\
WD 2034-532 & 20:38:16.84 & -53:04:25.4 & 14.56 & 0.12 & -0.11 & NA & NA & NA & 14.533 & 0.014 & 14.190 & 0.007 \\
WD 2059+013 & 21:02:19.3 & +01:32:15.9 & 13.93 & 0.09 & -0.08 & 14.21 & 0.23 & -0.19 & 13.982 & 0.007 & 14.364 & 0.005 \\
WD 2110+127 & 21:13:21.06 & +12:57:9.4 & 12.41 & 0.12 & -0.11 & NA & NA & NA & 12.787 & 0.006 & 13.182 & 0.005 \\
BPS CS 22951-0067 & 21:49:38.91 & -43:06:14.28 & 14.70 & 0.08 & -0.08 & 16.35 & 0.53 & -0.36 & 13.511 & 0.010 & 14.312 & 0.004 \\
WD 2149+021 & 21:52:25.38 & +02:23:19.54 & 12.24 & 0.06 & -0.06 & 12.51 & 0.12 & -0.11 & 12.604 & 0.007 & 12.688 & 0.005 \\
WD 2207-303 & 22:10:29.22 & -30:05:43.7 & 12.63 & 0.06 & -0.06 & NA & NA & NA & 12.596 & 0.004 & 13.539 & 0.003 \\
HS 2237+8154 & 22:37:15.57 & +82:10:27.32 & 17.45 & 1.40 & -0.59 & NA & NA & NA & 17.396 & 0.040 & 16.923 & 0.021 \\
WD 2246+223 & 22:49:5.76 & +22:36:32.31 & 17.57 & 1.27 & -0.57 & NA & NA & NA & 17.736 & 0.038 & 15.109 & 0.006 \\
WD 2316+123 & 23:18:45.1 & +12:36:2.9 & 16.70 & 0.56 & -0.37 & 15.74 & 0.26 & -0.21 & 16.889 & 0.045 & 15.783 & 0.011 \\
WD 2331-475 & 23:34:2.2 & -47:14:26.5 & 12.17 & 0.08 & -0.07 & NA & NA & NA & 12.419 & 0.004 & 13.405 & 0.003 \\
WD 2333-002 & 23:35:41.47 & +00:02:19 & 14.22 & 0.12 & -0.11 & NA & NA & NA & 14.149 & 0.014 & 14.719 & 0.010 \\
WD 2342+806 & 23:45:2.26 & +80:56:59.7 & 12.49 & 0.07 & -0.07 & 13.27 & 0.11 & -0.10 & 12.954 & 0.005 & 13.520 & 0.004 \\
WD 2353+026 & 23:56:27.72 & +02:57:5.61 & 13.85 & 0.11 & -0.10 & 14.58 & 0.24 & -0.20 & 13.829 & 0.007 & 14.357 & 0.006 \\

\hline

\end{tabular}
  \label{tab:wd-iue_and_galex}
\tablefoot{ NA means no available data.}
\end{scriptsize}
\end{table*}

As shown in  Figure~2, GALEX and IUE based UV photometries compare well except for 
the {\it brightest} sources; GALEX photometry is affected by photon count loses at high count rates. 
Following Camarota \& Holberg (2014, hereafter CH), we have fitted the samples to a
quadratic function  using the least-squares method:
\noindent
\begin{equation}
M_{GAL} = c_{0} + c_{1}M_{IUE} + c_{2}M_{IUE}^2
\end{equation}
\noindent
with $M_{GAL}$ the WD magnitude as per the GALEX AIS catalogue and $M_{IUE}$, the IUE-based synthetic
magnitude derived in this work. The coefficients of the fit, $c_{0}$, $c_{1}$ and $c_{2}$,
are given in Table ~ 3.  We have found a very good agreement with CH’s fits in the FUV photometry 
and a significant discrepancy in the NUV band that we ascribe to a possible typographic error 
in the parameters in CH’s Table~2. The FUV, NUV synthetic photometry for the rest of the WDs  in the IUE sample is provided in Table~4.

\begin{figure}[h!]
  \includegraphics[width = 12 cm]{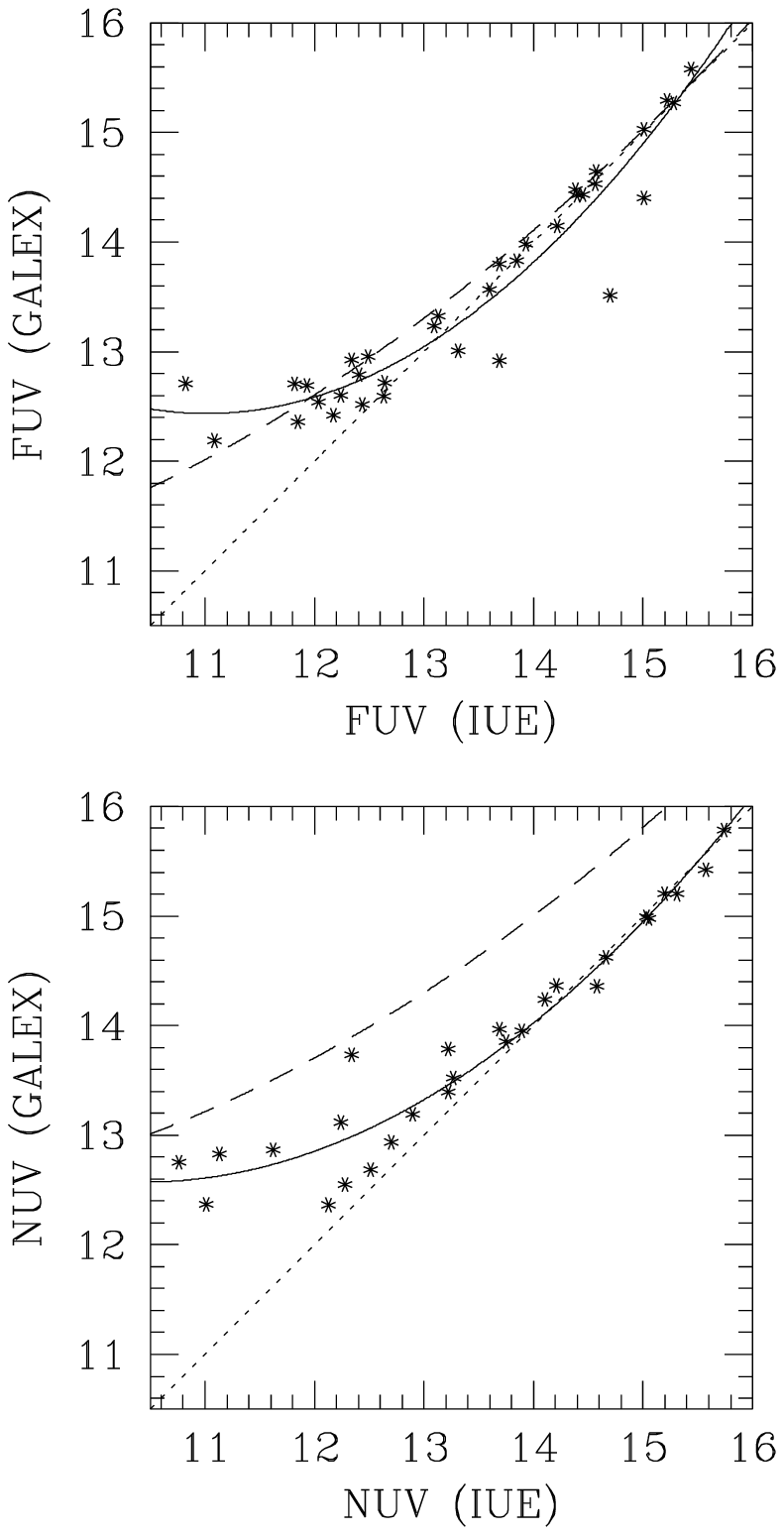}
  \caption{IUE versus GALEX photometry for the WDs sample. The dotted line represents the 1:1 correspondence, the
solid line the fit in Table~\ref{tab:fits} and the dashed line CH's fit.}

  \label{fig:fits}
\end{figure}

\begin{table}[h!]
 \caption{Quadratic fit parameters for the WDs sample}
 \centering
 \begin{tabular}{lll}
 \hline \hline
 Property & FUV & FUV\\
 \hline
c$_0$ & 31.2738  & 24.9204 \\
c$_1$ &-3.4197 & -2.3688 \\
c$_2$ & 0.1552 & 0.1136 \\
$\chi^2/dof$ &  0.0064& 0.0058  \\
No. of stars & 34 & 24 \\
Lower bound& 10.8 &10.8\\
Upper bound& 15.5 &15.7 \\
\hline
\end{tabular}
\label{tab:fits}
\end{table}
\begin{table*}[]
  \centering
  \caption{Synthetic photometry in the GALEX bands of WD observed with IUE: WDs without GALEX counterpart} 
  \begin{footnotesize}
  \begin{tabular}{lllllllll}
    
    \hline
    ID IUE & RAS (ICRS) & DEC (ICRS) & \multicolumn{4}{c}{IUE based synthetic photometry}  \\
              &                   &                  &  FUV & +eFUV & -eFUV & NUV & +eNUV & -eNUV  \\
    \hline

WD 2357+296 & 00:00:7.25 & +29:57:0.31 & 13.06 & 0.11 & -0.10 & NA & NA & NA  \\
WD 0017+136 & 00:20:1.79 & +13:52:48.1 & 15.27 & 0.12 & -0.10 & 15.09 & 0.17 & -0.14  \\
WD 0101+039 & 01:04:21.68 & +04:13:37.06 & 10.89 & 0.11 & -0.10 & 11.23 & 0.12 & -0.10  \\
WD 0104+50 & 01:07:11.02 & +51:10:22.74 & 11.85 & 0.06 & -0.06 & NA & NA & NA  \\
WD 0128-387 & 01:30:28.03 & -38:30:38.7 & 15.84 & 0.26 & -0.21 & 15.37 & 0.16 & -0.14  \\
WD 0131-164 & 01:34:24.07 & -16:07:8.38 & 11.98 & 0.06 & -0.06 & 12.83 & 0.11 & -0.10  \\
WD 0135-052 & 01:37:59.34 & -04:59:44.3 & NA & NA & NA & 15.15 & 0.26 & -0.21  \\
WD 0141-675 & 01:43:1.04 & -67:18:29.37 & NA & NA & NA & 17.13 & 3.06 & -0.72  \\
WD 0148+467 & 01:52:2.96 & +47:00:6.65 & 12.60 & 0.06 & -0.05 & 12.67 & 0.14 & -0.12  \\
WD 0214+568 & 02:17:33.52 & +57:06:47.5 & 12.49 & 0.07 & -0.06 & NA & NA & NA  \\
PG 0216+032 & 02:19:19 & +03:26:54 & 12.45 & 0.06 & -0.05 & 13.18 & 0.13 & -0.11  \\
WD 0227+050 & 02:30:16.62 & +05:15:50.68 & 12.10 & 0.06 & -0.05 & 12.43 & 0.10 & -0.09  \\
WD 0229-481 & 02:30:53.3 & -47:55:26.2 & 12.41 & 0.06 & -0.06 & NA & NA & NA  \\
WD 0232+525 & 02:36:19.54 & +52:44:12.5 & 13.33 & 0.07 & -0.06 & NA & NA & NA  \\
KUV 02503-0238 & 02:52:51 & -02:25:17.99 & 14.63 & 0.12 & -0.11 & 14.76 & 0.52 & -0.35  \\
WD 0255-705 & 02:56:17.21 & -70:22:10.86 & NA & NA & NA & 14.84 & 1.08 & -0.53  \\
WD 0346-011 & 03:48:50.19 & -00:58:32.02 & 12.02 & 0.11 & -0.10 & 12.73 & 0.09 & -0.08  \\
KUV 685-13 & 04:50:13.52 & +17:42:6.21 & 12.07 & 0.09 & -0.08 & 12.59 & 0.16 & -0.14  \\
WD 0455-282 & 04:57:13.9 & -28:07:54 & 11.76 & 0.07 & -0.06 & 12.41 & 0.17 & -0.15  \\
WD 0509-007 & 05:12:6.39 & -00:42:6 & 12.01 & 0.06 & -0.06 & NA & NA & NA  \\
WD 0531-022 & 05:34:18 & -02:15:0 & 14.28 & 0.43 & -0.31 & NA & NA & NA  \\
WD 0612+177 & 06:15:18.69 & +17:43:41 & 11.93 & 0.07 & -0.07 & 12.48 & 0.17 & -0.15  \\
WD 0640+015 & 06:43:16.02 & +01:30:12.6 & 13.81 & 0.08 & -0.08 & NA & NA & NA  \\
WD 0651-020 & 06:54:13 & -02:09:12 & 12.94 & 0.07 & -0.06 & 13.59 & 0.21 & -0.17  \\
WD 0715-703 & 07:15:16.59 & -70:25:5.6 & 12.08 & 0.07 & -0.06 & NA & NA & NA  \\
WD 0836+237 & 08:39:33.3 & +23:34:9 & 14.43 & 0.45 & -0.32 & NA & NA & NA  \\
BD+48 1777 & 09:30:46.78 & +48:16:23.77 & 8.87 & 0.05 & -0.05 & 9.29 & 0.12 & -0.11  \\
PG 0934+553 & 09:38:20.35 & +55:05:50.08 & 10.47 & 0.07 & -0.07 & NA & NA & NA  \\
WD 1013-050 & 10:16:28.68 & -05:20:32.06 & 12.07 & 0.06 & -0.06 & 12.71 & 0.11 & -0.10  \\
WD 1042-690 & 10:44:10.23 & -69:18:18.03 & 11.88 & 0.05 & -0.05 & 12.29 & 0.10 & -0.09  \\
WD 1105-048 & 11:07:59.95 & -05:09:25.89 & 12.90 & 0.06 & -0.05 & 13.08 & 0.10 & -0.10  \\
WD 1123+189 & 11:26:19.06 & +18:39:17.85 & 12.08 & 0.06 & -0.05 & 12.73 & 0.10 & -0.09  \\
WD 1234+482 & 12:36:45.18 & +47:55:22.34 & 12.30 & 0.09 & -0.08 & 12.94 & 0.11 & -0.10  \\
WD 1302+597 & 13:04:32.19 & +59:27:32.78 & 13.33 & 0.07 & -0.07 & 13.66 & 0.16 & -0.14  \\
WD 1321+36 & 13:23:35.26 & +36:07:59.51 & 10.11 & 0.05 & -0.05 & 10.40 & 0.10 & -0.09  \\
BD-07 3632 & 13:30:13.64 & -08:34:29.5 & 12.33 & 0.10 & -0.09 & 12.40 & 0.18 & -0.15  \\
WD 1337+705 & 13:38:50.48 & +70:17:7.66 & 11.84 & 0.06 & -0.05 & 12.23 & 0.12 & -0.11  \\
WD 1403-077 & 14:06:4.83 & -07:58:31.21 & 13.66 & 0.08 & -0.08 & NA & NA & NA  \\
WD 1424+534 & 14:25:55.46 & +53:15:25.14 & 13.50 & 0.07 & -0.07 & 14.23 & 0.21 & -0.18  \\
CD-37 10500B & 15:47:30.07 & -37:55:8.11 & 16.76 & 1.73 & -0.64 & 13.97 & 0.20 & -0.17  \\
WD 1615-154 & 16:17:55.26 & -15:35:51.93 & 11.65 & 0.08 & -0.08 & 12.28 & 0.11 & -0.10  \\
CD-38 10980 & 16:23:33.84 & -39:13:46.16 & 9.62 & 0.06 & -0.06 & 10.15 & 0.10 & -0.09  \\
WD 1639+537 & 16:40:57.16 & +53:41:9.6 & NA & NA & NA & 17.09 & 0.86 & -0.48  \\
WD 1657+343 & 16:58:51.12 & +34:18:53.29 & 14.37 & 0.10 & -0.09 & NA & NA & NA  \\
WD 1713+695 & 17:13:6.12 & +69:31:25.7 & 13.07 & 0.15 & -0.13 & 13.13 & 0.51 & -0.34  \\
WD 1725+586 & 17:26:43.36 & +58:37:32.06 & 13.41 & 0.25 & -0.20 & NA & NA & NA  \\
WD 1900+706 & 19:00:10.25 & +70:39:51.2 & 13.88 & 0.15 & -0.13 & NA & NA & NA  \\
** LDS 678A & 19:20:34.92 & -07:40:0.07 & 14.83 & 0.23 & -0.19 & NA & NA & NA  \\
WD 1936+327 & 19:38:28.21 & +32:53:19.9 & 12.54 & 0.08 & -0.08 & NA & NA & NA  \\
LS II +18 9 & 19:43:31.21 & +18:24:34.58 & 10.20 & 0.07 & -0.06 & 10.77 & 0.10 & -0.09  \\
WD 2007-303 & 20:10:56.85 & -30:13:6.64 & 12.13 & 0.06 & -0.06 & 12.28 & 0.11 & -0.10  \\
RX J2013.1+4002 & 20:13:9.37 & +40:02:24.25 & 12.10 & 0.06 & -0.06 & NA & NA & NA  \\
WD 2028+390 & 20:29:56.16 & +39:13:32 & 12.02 & 0.06 & -0.06 & 12.56 & 0.17 & -0.14  \\
WD 2123-82 & 21:31:5.18 & -82:40:53.25 & 12.46 & 0.07 & -0.06 & 12.35 & 0.20 & -0.17  \\
WD 2211-495 & 22:14:11.91 & -49:19:27.26 & 9.40 & 0.05 & -0.05 & 10.23 & 0.10 & -0.09  \\
WD 2313-021 & 23:16:12.42 & -01:50:35.06 & 11.79 & 0.07 & -0.06 & 12.09 & 0.10 & -0.09  \\
GD 1110 & 23:19:24.43 & -08:52:37.91 & 11.87 & 0.06 & -0.05 & NA & NA & NA  \\
WD 2317-054 & 23:19:58.4 & -05:09:56.16 & 10.07 & 0.05 & -0.05 & 10.47 & 0.10 & -0.09  \\
GD 1309 & 23:29:12 & -10:05:0 & 11.68 & 0.07 & -0.06 & 12.09 & 0.11 & -0.10  \\
WD 2349+286 & 23:51:56 & +28:55:12 & 14.39 & 0.22 & -0.18 & NA & NA & NA  \\
 \hline

\end{tabular}
  \label{tab:wd-iue}
\tablefoot{ NA means no available data.}
\end{footnotesize}
\end{table*}

\begin{figure*}[]
  \includegraphics[width=15cm]{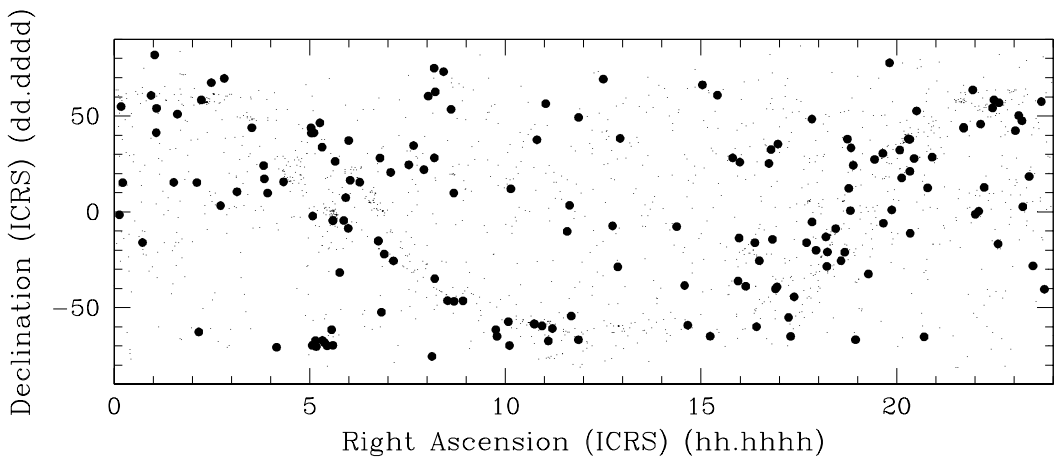}
  \caption{Distribution in the sky of the sources in the catalogue. Variable sources
(as per the criteria in Sect.~4.4) are indicated with filled circles.}

  \label{fig:skymap}
\end{figure*}

\section{The catalogue}

The catalogue contains FUV magnitudes for all stars (with and without multiple observations) and
NUV magnitudes only for non variable stars, or stars with just one good observation, as pointed out
in Sect.~4.2. The number of observations used for the variability evaluation
are indicated for all catalogue entries; $\sim 65$\% of the sources have only one good observation. 

Appendix~A contains an excerpt of the on-line catalogue. For each source, the following entries are provided:
\begin{itemize}
\item Object identification in the IUE Archive.
\item Right Ascension and Declination (ICRS).
\item No. of SWP observations used to compute FUV. 
\item Synthetic FUV magnitude (Eq. 4) computed from the mean flux, if several observations are used (see section 4.2). 
\item Error in the synthetic FUV magnitude  computed from the mean flux and error (see section 4.3). 
\item No. of LW observations used to compute NUV. 
\item Synthetic NUV magnitude (obtained from SWP and LW mean fluxes).
\item Error in the synthetic NUV magnitude (obtained from SWP and LW mean fluxes).
\end{itemize}

 For the 89 stars found to be variable in the SW range, additional entries are provided with the  FUV synthetic magnitude for each
observation  (see Sect.~4.4);  the following entries are provided:
\begin{itemize}
\item Object identification in the IUE Archive.
\item Right Ascension and Declination (ICRS).
\item Observation date and time.
\item Synthetic FUV magnitude computed from the flux.
\item Error in the synthetic FUV magnitude  computed from the flux and error. 
\end{itemize}

(see an excerpt in Appendix~B).

The catalogue contains synthetic FUV magnitudes for  1,631 sources, ranging from FUV = 1.81 to FUV = 18.65. 
In the NUV band, the catalogue includes observations for 
1,005 stars ranging from NUV = 3.08 to NUV= 17.74 mag. 
The distribution of sources in the sky is plotted in Figure~3; notice the good coverage of the Galactic plane.
A summary statistics of the catalogue contents is available in Table~4. This 
work adds UV photometry for  1,493 new sources, most of them hot (O-A spectral type) stars.
The catalogue is available to the community through the services of the Centre de Donn\'ees Stellaires.

\begin{table}
 \caption{Catalogue contents}
 \centering
\begin{scriptsize}
 \begin{tabular}{llcccc}
 \hline \hline
\multicolumn{2}{c}{IUE Class} &\multicolumn{2}{c}{Number of Stars} & \multicolumn{2}{c}{No. of IUE spectra used}\\
ID&   Description & All & in GALEX$^{(a)}$& SW & LW \\
 \hline
10 & WC & 22 & 0 & 70 & 57 \\
11 & WN & 38 & 0 & 165 & 49 \\
12 & Main Sequence O & 106 & 1  & 150 & 120 \\
13 & Supergiant O & 38 & 0 & 58 & 27 \\
14 & Oe & 4 & 0 & 6 & 2 \\
15 & Of & 10 & 0 & 19 & 5 \\
16 & SD O & 68 & 29 & 176 & 457 \\
17 & WD O & 15 & 5 & 20 & 62 \\
20 & B0-B2 V-IV & 143 & 5 & 294 & 294 \\
21 & B3-B5 V-IV & 68 & 1 & 87 & 110 \\
22 & B6-B9.5 V-IV & 139 & 7 & 218 & 139 \\
23 & B0-B2 III-I & 111 & 2 & 169 & 64 \\
24 & B3-B5 III-I & 25 & 1 & 48 & 24 \\
25 & B6-B9.5 III-I & 57 & 3 & 85 & 43 \\
26 & Be & 29 & 1 & 68 & 93 \\
27 & Bp & 36 & 2 & 63 &  57 \\
28 & sd B & 36 & 15 & 39 & 27 \\
29 & WDB & 13 & 7 & 46 & 32 \\
30 & A0-A3 V-IV & 111 & 5 & 150 &  117\\
31 & A4-A9 V-IV & 28 & 2 & 32 & 21 \\
32 & A0-A3 III-I & 25 & 2 & 31 & 17 \\
33 & A4-A9 III-I & 17 & 2 & 20 & 8 \\
34 & Ae & 4 & 1 & 55 & 4 \\
35 & Am & 13 & 1 & 16 & 12 \\
36 & Ap & 20 & 1 & 70 & 87 \\
37 & WDA & 73 & 26 & 101 & 102 \\
38 & Horizontal Branch Stars & 26 & 6 & 33 & 18 \\
40 & F0-F2 & 39 & 3 & 41 & 51 \\
41 & F3-F9 & 52 & 8 & 60 &  47 \\
42 & Fp & 1 & 1 & 1 & 1 \\
44 & G IV-V & 74 & 20 & 210 & 106 \\
45 & G III-I & 40 & 7 & 66 & 37 \\
46 & K V-IV & 50 & 14 & 323 &  112 \\
47 & K III-I & 35 & 8 & 61 & 16 \\
48 & M V-IV & 22 & 2 & 106 & 52 \\
49 & M III-I & 15 & 5 & 35 & 16 \\
50 & R, N or S Types & 7 & 4 & 9 & 6 \\
51 & Long Period Variable Stars & 1 & 0 & 1 & 3 \\
52 & Irregular Variables & 10 & 0 & 18 & 15 \\
53 & Regular Variables & 30 & 10 & 24 & 104 \\
54 & Dwarf Novae & 33 & 17 & 62 & 61 \\
55 & Classical Novae & 21 & 3 & 39 & 51 \\
58 & T Tauri & 20 & 5 & 30 & 96 \\
\hline
\end{tabular}
\begin{flushleft}
\begin{tabular}{ll}
(a) & Identified by cross correlation with the GALEX AIS GR5 catalogue. A search \\
 & radius of 3 arcsec is used.\\
\end{tabular}
\end{flushleft}
\end{scriptsize}
\label{tab:summary}
\end{table}

\section{Conclusions}
 
From an initial sample of 31,982 stellar IUE spectra, we have computed the synthetic photometry for:
\begin{itemize}
\item 1,631 sources in the GALEX FUV band with magnitudes ranging from FUV = 1.81 to FUV = 18.65.
\item  1,005 sources in the GALEX NUV band with NUV ranging from NUV = 3.08 to NUV= 17.74 mag. 
\end{itemize}

The  FUV and NUV synthetic photometry compares well with GALEX. A sample of WD’s observed with IUE and GALEX has been used for the test;
a good agreement with CH has been found for the FUV band but no for the NUV band.

The catalogue is available to the community 
through the services of the Centre de Donn\'ees Stellaires. It
adds UV photometry for 1,493 new sources with respect to the GALEX AIS catalogue, most of them hot (O-A spectral type) stars.
The sources in the catalogue are distributed over the full sky, including the Galactic plane.
  
\Online

\begin{appendix}
\section{Catalogue excerpt}

An excerpt (first 10 entries) of the on-line catalogue is shown in Table~A.1 (see Sect.~6, for a detailed description of the fields).
\begin{table*}[h!]
\scriptsize
\begin{tabular}{ccccccccccc}
\hline
Object & RA(2000)& Dec(2000) & NobsSWP & FUV & +eFUV &-eFUV & NobsLW & NUV& +eNUV &-eNUV\\
          & (hh:mm:ss.ss) & ($\pm$dd:mm:ss.ss) & &(ABmag) & (ABmag) & (ABmag) & & (ABmag) & (ABmag) & (ABmag) \\
\hline
CD-40 15307 & 00:00:20.14 & -39:23:55.24 & 1 & 10.14 & 0.05 & -0.05 & 1 & 10.16 & 0.11 & -0.10 \\
WD 2357+296 & 00:00:7.25 & +29:57:0.31 & 2 & 13.06 & 0.11 & -0.10 & 0 & NA & NA & NA \\
HD 225094 & 00:03:25.71 & +63:38:25.88 & 3 & 8.81 & 0.04 & -0.04 & 1 & 8.64 & 0.08 & -0.07 \\
HD 225132 & 00:03:44.39 & -17:20:9.57 & 6 & 6.42 & 0.04 & -0.04 & 1 & 6.08 & 0.05 & -0.04 \\
HD 108 & 00:06:3.39 & +63:40:46.76 & 1 & 8.36 & 0.05 & -0.05 & 0 & NA & NA & NA \\
HD 186 & 00:06:47.96 & +44:36:46.2 & 1 & 9.07 & 0.05 & -0.05 & 1 & 9.21 & 0.11 & -0.10 \\
BD+59 2829 & 00:06:48.3 & +60:36:0.83 & 1 & 11.85 & 0.05 & -0.05 & 0 & NA & NA & NA \\
PG 0004+133 & 00:07:33.78 & +13:35:57.66 & 1 & 13.04 & 0.09 & -0.08 & 1 & 13.28 & 0.17 & -0.15 \\
WD 0005+511 & 00:08:18.17 & +51:23:16.6 & 1 & 11.09 & 0.06 & -0.06 & 0 & NA & NA & NA \\
HD 358 & 00:08:23.26 & +29:05:25.55 & 1 & 3.19 & 0.05 & -0.05 & 0 & NA & NA & NA \\
\hline
\end{tabular}
  \caption{Catalogue layout (10 first entries).}
  \label{}
\end{table*}

\section{FUV magnitude - Variable stars}
An excerpt (first 3 entries) of the list of FUV magnitudes for variable
stars is shown in Table B.1 (see Sect.~6 for a  description of the fields).  
\begin{table*}[h!]
\begin{tabular}{cccccccc}
\hline
Object & RA(ICRS) & Dec(ICRS) & Obsdate & Obstime & FUV & +eFUV & -eFUV \\
\hline
HD 352 & 00:08:12.1 & -02:26:51.76 & 1985-08-02 & 10:10:00 & 15.02 & 0.19 & -0.16 \\
HD 352 & 00:08:12.1 & -02:26:51.76 & 1981-05-20 & 02:02:03 & 14.34 & 0.08 & -0.08 \\
HD 352 & 00:08:12.1 & -02:26:51.76 & 1984-06-14 & 02:06:56 & 15.19 & 0.15 & -0.13 \\
HD 5394 & 00:56:42.53 & +60:43:0.27 & 1982-01-28 & 00:37:38 & 6.57 & 0.53 & -0.35 \\
HD 5394 & 00:56:42.53 & +60:43:0.27 & 1988-07-10 & 13:18:22 & 1.90 & 0.03 & -0.03 \\
HD 5394 & 00:56:42.53 & +60:43:0.27 & 1988-07-10 & 14:46:38 & 1.84 & 0.03 & -0.03 \\
HD 5679	&01:02:18.45 &+81:52:32.08	& 1978-05-15	& 17:07:54 &	9.46& 0.06&	-0.05\\
HD 5679	&01:02:18.45 &+81:52:32.08	& 1981-08-04	& 08:17:42 &	15.11&0.68& -0.41\\
\hline
 \end{tabular}
  \caption{FUV magnitude for variable stars layout (10 first entries).}
  \label{vstars}
\end{table*}

\end{appendix}


\begin{thebibliography}{10}

\bibitem[Bianchi et al.(2011)]{bianchi:11}
Bianchi, L.,  Herald, J., Efremova, B., et al., 2011, ApSS, 335, 161

\bibitem[Bianchi (2014)]{bianchi:14}
Bianchi, L., 2014, ApSS, 304, 103

\bibitem[Boggess et al (1978)]{boggess:78}
Boggess, A., Carr, F.A., Fischel, D. et al., 1978, Nature, 275, 372

\bibitem[Bohlin and Holm et al.(1980)]{bohlin:80}
Bohlin, R.C., Holm, A.V., Savage, D.V. et al.,  1980, Astronomy \& Astrophysics, 85, 1

\bibitem[Camarota and Holberg(2014)]{camarotaholberg:14}
Camarota, L., Holberg, J.B. 2014, MNRAS, 438, 3111

\bibitem[Martin et al.(2005)]{martin:05}
Martin, D.C., Fanson, J.,  Schiminovich, D. et al., 2005, ApJ, 619, L1

\bibitem[Morrissey et al.(2005)]{Morrissey:05}
Morrissey, P., Schiminovich, D., Barlow, T. A. et al., 2005, ApJ, 619, L7

\bibitem[Morrissey et al.(2007)]{Morrissey:07}
Morrissey, P., Conrow, T., Barlow, T. A. et al., 2007, ApJS, 173, 682 

\bibitem[Rodr\'iguez-Pascual et al.(2007)]{rod-pas:99}
Rodr\'iguez-Pascual, P., Gonzalez-Riestra, R., Schartel, N.  et al., 1999, A\&AS, 139, 183

\end{thebibliography}

\section*{Acknowledgements}
This work has ben supported by the Ministry of Economy and Competitiveness of Spain through grants: AYA2011-29754-c3-01,
ESP2014-54243-R.
Leire Beitia-Antero acknowledges the receipt of a "Beca de Colaboraci\'on” from the Ministry of Education of Spain.

\end{document}